\documentclass[hyper,11pt,letterpaper]{JHEP3}

\usepackage{cite, graphicx, subfigure, amsmath,amssymb,calrsfs,mathrsfs} 

\newcommand{\rv}[1]{\boldsymbol{\mathsf{#1}}}
\newcommand{\ra}{\rightarrow}

\newcommand{\vev}[1]{\left\langle #1\right\rangle}

\title{Universal, Continuous-Discrete Nonlinear Yau Filtering I: Affine, Linear State Model with State-Independent Diffusion Matrix}

\author{Bhashyam Balaji\\
Radar Systems Section,\\
Defence Research and Development Canada, Ottawa,\\
3701 Carling Avenue, \\
Ottawa ON K1A 0Z4 Canada\\
Email: Bhashyam.Balaji@drdc-rddc.gc.ca\\
}

\abstract{
The continuous-discrete filtering problem requires the solution of a partial differential equation known
as the Fokker-Planck-Kolmogorov forward equation (FPKfe). In this paper, it is pointed out that for a state model with an affine, linear drift and state-independent diffusion matrix the fundamental solution can be obtained using only linear algebra techniques. In particular, no differential equations need to be solved. Furthermore, there are no restrictions on the size of the time step size, or on the measurement model. Also discussed are important computational aspects that are crucial for potential real-time implementation for higher-dimensional problems. The solution is universal in the sense that the initial distribution may be arbitrary.}


\begin{document}

\section{Introduction}

The problem of continuous-discrete (continuous-continuous) filtering is to estimate the state that is described by a continuous-time stochastic process from the observations of a related discrete-time (continuous-time) stochastic process called the measurement process. The complete solution of the filtering problem, in the Bayesian sense, is given by the conditional probability density function. Given the conditional probability density we can define optimality under any criteria. Usually, the conditional mean, which is the minimum mean-squares estimate, is studied. A solution is said to be universal if the initial distribution is arbitrary (see, for example, \cite{A.H.Jazwinski1970}).

Continuous-discrete filtering requires the solution of a partial differential equation known as the Fokker-Planck-Kolmogorov forward equation (FPKfe) \cite{H.Risken1999}. In general, such problems are often solved via discretization of the PDE, such as using the method of finite differences, \cite{Thom'ee90,Marchuk90}, spectral methods \cite{C.CanutoandM.Y.HussainiandA.QuarteroniandT.A.ZangJr.2006} and finite element methods\cite{V.Thom'ee1997}. The numerical solution of a PDE is non-trivial. In particular, there are severe time step size and grid spacing restrictions (especially for high accuracy solutions) for stability, accuracy, and consistency of the numerical method. In addition, a plain implementation places very large computational and memory requirements, especially for higher dimensional problems. It is therefore desirable to be able to exactly solve for the fundamental solution of the FPKfe, in terms of which the FPKfe can be solved.  

In this paper it is pointed out that the fundamental solution of the FPKfe for the affine, linear state model with state-independent diffusion matrix, a special case of the Yau filter (discussed below), can be computed using only linear algebra methods. In particular, the fundamental solution for such a model can be computed exactly for an \textit{arbitrary time step size} without having to solve a differential equation. It is also noted that the computational requirements are considerably less than that suggested by the size of the grid space.  In particular, the conditional probability density can be computed efficiently (especially, concerning memory requirements) and rapidly with  the use of sparse multilinear data structures. In many applications, such as target tracking, the state model is linear and the measurement model is nonlinear. Therefore, the results in the paper are applicable to a large number of practical problems. 

The Yau filter is defined to be one where state model drift a linear plus a gradient term satisfying the Ben\v es condition and with a time- and state- independent diffusion matrix \cite{S.-T.Yau2003}. In continous-continuous filtering theory, the Yau filtering system is defined as one with an additional restriction that the measurement process is a continuous-time stochastic process with linear drift. The Yau filtering system plays a fundamental role in continuous-continuous filtering. The reason is that it has been shown that a finite-dimensional filter of maximal rank has to be a Yau filter\cite{S.-T.Yau2003}. In light of its importance, some real-time implementable solutions for the Yau filtering system have been investigated. In \cite{S.-T.YauLai2003,B.Balaji2006}, the solution of the Yau equation has been shown to be equivalent to a solution of a system of linear ODEs. The case of nonlinear observations has also been investigated in some papers. In particular, it is shown in  \cite{YauYau2005} that some nonlinear observation models can also be solved via ODEs. Finally, in \cite{YauS-T.Yau2004}, a solution is presented in terms of a power series. 

In this paper, the special case of the continuous-discrete Yau filter, namely one with an affine, linear state model drift and state-independent (but possibly time-dependent) diffusion matrix, is studied. Note that the measurement model in the continuous-discrete Yau case can be arbitrary, and so a more general filtering problem is solved than for the continuous-continuous Yau case.  In a following paper, it is shown that the fundamental solution of the FPKfe for another special case of the Yau filter, namely one with a gradient drift satisfying a certain condition (and also known as the Ben\v es model drift),  can also be solved exactly using such elementary methods.  

It is noted that the methods developed for solving the continuous-continuous filtering problem quoted in the papers above can be used to solve the prediction part of the continuous-discrete filtering problem. However, they require solution of ODEs, or a power series expression, rather than the closed form expressions derived here. 

The relevant aspects of filtering theory are reviewed in Section \ref{sec:RevCDFilt}. The fundamental solutions are derived for the additive noise case in Section \ref{sec:LangFundSol} and \ref{sec:FundSolConsDiff}. Some implementational aspects are discussed in Section \ref{sec:PractImpl} and related comments on computational aspects of discrete-discrete filtering are presented in  Section \ref{sec:Disc-DiscFilt}. An example is presented in Section \ref{sec:Examples}.  In Appendix \ref{sec:GenTimeDep}, the formula for the state transition matrix for the general time-dependent case is reviewed. In Appendix \ref{sec:AppManTargets}, the linear manoeuvering target tracking models reviewed in \cite{X.R.LiV.P.Jilkov2003} are explicitly solved. Some of the results are well-known and have been derived elsewhere, sometimes presented in a different form. Here use is made of the elegant method of \cite{ChengYau1September1997} that enables a simple and systematic derivation of explicit formulas for arbitrary affine, linear models up to four-dimensional state space problems.


\section{Review of Continuous-Discrete Filtering Theory}\label{sec:RevCDFilt}

\subsection{Langevin Equation, the FPKfe and its Fundamental Solution}

  The general continuous-time state model is described by the following stochastic differential equation (SDE):
\begin{align}\label{eq:LangevinEquation}
	d\rv{x}(t)=f(\rv{x}(t),t)dt+e(\rv{x}(t),t)d\rv{v}(t),\qquad \rv{x}(t_0)=x_0.
\end{align}
Here $\rv{x}(t)$ and $f(\rv{x}(t), t)$ are $n-$dimensional column vectors, the diffusion vielbein $e(\rv{x}(t), t)$ is an $n\times p$ matrix and $\rv{v}(t)$ is a $p-$dimensional column vector. The noise process $\rv{v}(t)$ is assumed to be Brownian with covariance $Q(t)$ and the quantity $g\equiv eQe^T$ is termed the diffusion matrix. All functions are assumed to be sufficiently smooth. Equation \ref{eq:LangevinEquation} is also referred to as the Langevin equation. It is interpreted in the It\^o sense.

Let $\sigma_0(x)$ be the initial probability distribution of the state process. Then, the evolution of the probability distribution of the state process described the Langevin equation, $p(t, x)$, is described by the FPKfe, i.e.,
\begin{align}
        {\left\lbrace\begin{aligned}    
                \frac{\partial p}{\partial t}(t,x)&=-\sum_{i=1}^n\frac{\partial}{\partial x_i}\left[ f_i(x,t)p(t,x) \right]+\frac{1}{2}\sum_{i,j=1}^n\frac{\partial^2}{\partial x_i\partial x_j}\left[g_{ij}(t,x)p(t,x)\right], \\ 
        p(t_0,x)&=\sigma_0(x).
\end{aligned}
\right.}
\end{align}

Let $t'' > t'$, and consider the following PDE:
\begin{align}
        {\left\lbrace\begin{aligned}    
                \frac{\partial P}{\partial t}(t,x|t',x')&=-\sum_{i=1}^n\frac{\partial}{\partial x_i}\left[ f_i(x,t)P(t,x|t',x') \right]+\frac{1}{2}\sum_{i,j=1}^n\frac{\partial^2}{\partial x_i\partial x_j}\left[g_{ij}(t,x)P(t,x|t',x'))\right], \\ 
        P(t',x''|t',x')&=\delta^n(x-x_0).\end{aligned}
\right.}
\end{align}
The quantity $P(t,x|t',x')$ is known as the fundamental solution of the FPKfe. In this instance, the physical interpretation is that it is the transition probability density. 

From the fundamental solution one can compute the probability at a later time for an arbitrary initial condition as follows\footnote{In this paper, all integrals are assumed to be from $-\infty$ to $+\infty$, unless otherwise specified.}:
\begin{align}\label{eq:EvolFPKfeSolnIntegral}
	p(t'',x'')=\int P(t'',x''|t',x')p(t',x')\left\{ d^nx' \right\}.
\end{align}
 Therefore, in order to solve the FPKfe it is sufficient to solve for the transition probability density $P(t, x|t', x')$. Note that this solution is universal in the sense that the initial distribution can be arbitrary.

\subsection{Continuous-Discrete Filtering}

In ths paper, it is assumed that the measurement model is described by the following discrete-time stochastic process
\begin{align}
	\rv{y}(t_k)=h(\rv{x}(t_k),t_k,\rv{w}(t_k)),\qquad k=1,2,\ldots,\qquad t_k>t_0,
\end{align}
where $y(t)\in \mathbb{R}^{m\times1}, h\in \mathbb{R}^{m\times1}$, and the noise process $\rv{w}(t)$ is assumed to be a white noise process. Note that $\rv{y}(t_0)=0$. It is assumed that $p(y(t_k)|x(t_k))$ is known.

Then, the universal continuous-discrete filtering problem can be solved as follows. Let the initial distribution be $\sigma_0(x)$ and let the measurements be collected at time instants $t_1, t_2,\ldots, t_k,\ldots$. We use the notation $Y(\tau) = \{y(t_l) : t_0 < t_l\le \tau\}$. Prior to incorporating the measurements, the state evolves according to the FPKfe, i.e.,
\begin{align}
	p(t_1,x_1|Y(t_0))=\int P(t_1,x|t_0,x_0)p(t_0,x_0)\left\{ d^nx_0 \right\}.
\end{align}
This is the prediction step.

According to Bayes' rule
\begin{align}
	p(t_1,x|Y(t_0))=\frac{p(y(t_1)|x,Y(t_0))p(t_1,x|Y(t_0))}{p(y(t_1)|Y(t_0))}.
\end{align}
Now $p(t_1,x|Y(t_1))=p(t_1,x|y(t_1),Y(t_0))$ and since measurement noise is white it follows that $p(y(t_1)|x,Y(t_0))=p(y(t_1)|x)$. Hence, the `corrected' conditional density at time $t_1$ is 
\begin{align}
	p(t_1,x|Y(t_1))=\frac{p(y(t_1)|x)p(t_1,x|Y(t_0))}{\int p(y(t_1)|\xi)p(t_1,\xi|Y(t_0))\left\{ d^n\xi \right\}}.
\end{align}
This is then the initial condition of the FPKfe for the next prediction step which results in
\begin{align}
 {\left\lbrace\begin{aligned} 
	 p(t_2,x|Y(t_1))&=\int P(t_2,x|t_1,x_1)p(t_1,x_1|Y(t_1))\left\{ d^nx_0 \right\},\qquad\text{(Prediction Step)},\\ 
	 p(t_2,x|Y(t_2))&=\frac{p(y(t_2)|x)p(t_2,x|Y(t_1))}{\int p(y(t_2)|\xi)p(t_2,\xi|Y(t_1))\left\{ d^n\xi \right\}},\qquad\text{(Correction Step)},
 \end{aligned}\right.}
\end{align}
and so on.

\section{Fundamental Solution I: Additive Noise}\label{sec:LangFundSol}
In this section, the following general affine, linear state model with additive noise is considered: 
\begin{align}
	d\rv{x}(t)=( F(t)\rv{x}(t)+ l(t))dt+e(t)d\rv{v}_i(t),\qquad i=1,\ldots, n.
\end{align}
It is assumed that $F(t)$ commutes at different times, i.e.,  
\begin{align}
	\left[ F(t),F(t') \right]&\equiv F(t)F(t')-F(t')F(t),\\ \nonumber
	&=0.
\end{align}
The case when $F(t)$ does not commute with itself at different times is discussed in Appendix \ref{sec:GenTimeDep}. Also, the matrix notation is used here. 

The state model can be written as the following Langevin equation
\begin{align}\label{eq:LangForm}
	\frac{d\rv{x}}{dt}(t)&=\left( F(t)\rv{x}(t)+l(t) \right)+e(t)\frac{d\rv{v}}{dt}(t),\\ \nonumber
	&=( F(t)\rv{x}(t)+l(t))+e(t)\rv{\nu}(t),
\end{align}
where $\rv{\nu}(t)$ is $\delta-$correlated and Gaussian distributed as follows:
\begin{align}
	\vev{\rv{\nu}_i(t)}&=0,\\ \nonumber
	\vev{\rv{\nu}_i(t)\rv{\nu}_j(t')}&=Q_{ij}(t)\delta(t-t'),\qquad Q_{ij}(t)=Q_{ji}.
\end{align}

The formal solution of Equation \ref{eq:LangForm} is 
\begin{align}
	\rv{x}(t)=U(t,t_0)\rv{x}(t_0)+\int_{t_0}^tU(t,\tau)[l(\tau)+e(\tau)\rv{\nu}(\tau)]d\tau,
\end{align}
where the state transition matrix is the solution of the following equation:
\begin{align}\label{eq:GenTimDepModelP}
 {\left\lbrace\begin{aligned}    
	 \frac{dU}{dt}(t,t_0)&=F(t)U(t,t_0),\\ 
	 U(t_0,t_0)&=I.
\end{aligned}
\right.}
\end{align}
This can be verified as follows. The Leibniz rule is
\begin{align}\label{eq:LeibnizRule}
	\frac{\partial}{\partial z}\int_{a(z)}^{b(z)}\Phi(\xi,z)dx=\int_{a(z)}^{b(z)}\frac{\partial \Phi}{\partial z}(x,z)d\xi+\Phi(b(z),z)\frac{\partial b}{\partial z}(z)-\Phi(a(z),z)\frac{\partial a}{\partial z}(z),
\end{align} 
where $\Phi(\xi,z),a(z), b(z)$ are assumed to be sufficiently smooth functions of their arguments. From the Leibniz rule it therefore follows that 
\begin{align}
	\frac{d\rv{x}}{dx}(t)&=F(t)U(t,t_0)\rv{x}(t_0)+\int_{t_0}^tF(t)U(t,\tau)\left[ l(\tau)+e(\tau)\rv{\nu}(\tau) \right]d\tau++U(t,t)(l(t)+e(t)\rv{\nu}(t)),\\ \nonumber
	&=F(t)\left[ U(t,t_0)\rv{x}(t_0)+\int_{t_0}^tU(t,\tau)\left[ l(\tau)+e(\tau)\rv{\nu}(\tau) \right]d\tau \right]+l(t)+e(t)\rv{\nu}(t),\\ \nonumber
	&=\left( F(t)\rv{x}(t)+l(t) \right)+e(t)\rv{\nu}(t).
\end{align}

Also, $U(t,t_0)$ satisfies the semi-group property:
\begin{align}
	U(t_2,t_0)=U(t_2,t_1)U(t_1,t_0), 
\end{align}
which immediately implies that
\begin{align}
	U(t,t_0)=U^{-1}(t_0,t).
\end{align}
Since the matrix $F(t)$ commutes at all times
\begin{align}
	U(t,t_0)=e^{\int_{t_0}^tF(\tau)d\tau}.
\end{align}
If the matrix $F$ is time-independent
\begin{align}
	U(t,t_0)=e^{F(t-t_0)}.
\end{align}

Since a linear transformation of Gaussian variables is also Gaussian, $\rv{x}(t)$  is also a Gaussian random variable. Therefore, it is completely characterized by the mean vector and covariance matrix. 

The mean vector is 
\begin{align}
	\mu(t,t_0)&=\vev{\rv{x}(t)},\\ \nonumber
	&=U(t,t_0)x(t_0)+\tilde{l}(t),\qquad \tilde{l}(t)\equiv \int_{t_0}^tU(t,t_0)l(t)dt,
\end{align}
and the covariance matrix is 
\begin{align}
	\Sigma(t,t_0)&=\vev{\left[ \rv{x}(t)-\mu(t) \right]\left[ \rv{x}(t)-\mu(t) \right]^T},\\ \nonumber
	&=U(t,t_0)\sigma(t_0)U^T(t,t_0)+\int_{t_0}^t\int_{t_0}^tU(t,\tau_1)e(\tau_1)Q(\tau_1)\delta(\tau_1-\tau_2)e^T(\tau_2)U^T(t,\tau_2)d\tau_1d\tau_2,\\ \nonumber
	&=U(t,t_0)\sigma(t_0)U^T(t,t_0)+\int_{t_0}^tU(t,\tau)g(\tau)U^T(t,\tau)d\tau.
\end{align}
Combining the results for $\mu(t,t_0)$ and $\sigma(t,t_0)$, the fundamental solution is 
\begin{align}
	P(t,x|t_0,x_0)=\frac{1}{\sqrt{(2\pi)^n\det \sigma(t,t_0)}}\exp\left( -\frac{1}{2}\left[ x-\mu(t,t_0) \right]^T\left[ \Sigma^{-1}(t,t_0) \right]\left[ x-\mu(t,t_0) \right] \right).
\end{align}

Finally, observe that
\begin{align}
	\frac{d\Sigma}{dt}(t)&=F(t)\left[U(t,t_0)\Sigma(t_0)U^T(t,t_0)\right]+	\left[ U(t,t_0)\Sigma(t_0)U^T(t,t_0) \right]F^T(t)+\frac{d}{dt}\int_{t_0}^tU(t,\tau)g(\tau)U^T(t,\tau)d\tau,\\ \nonumber
	&=F(t)\left[U(t,t_0)\Sigma(t_0)U^T(t,t_0)\right]+\left[ U(t,t_0)\Sigma(t_0)U^T(t,t_0) \right]F^T(t)\\ \nonumber
	&\qquad +U(t,t)g(t)U(t,t)+F(t)\int_{t_0}^t\left[U(t,\tau)G(\tau)U^T(t,\tau)\right]d\tau+\int_{t_0}^t\left[U(t,\tau)g(\tau)U^T(t,\tau)\right]F^T(\tau)d\tau,\\ \nonumber
&=F(t)\Sigma(t)+\Sigma(t)F^T(t)+g(t),
\end{align}
where the Leibniz rule (Equation \ref{eq:LeibnizRule}) has been used in the second step. This result will be used in the following section.

\section{Fundamental Solution II: State-Independent Diffusion Matrix}\label{sec:FundSolConsDiff}

In Section \ref{sec:LangFundSol}, the fundamental solution was derived for the time-independent affine, linear state model with additive noise. In this section, it is pointed out that a similar result follows if the noise is multiplicative but with state dependent diffusion matrix. This is a straightforward generalization of the result derived in \cite{WangUhlenbeck1945}; it also assumes that the diffusion matrix is time-independent.

The state model considered here is
\begin{align}
	d\rv{x}(t)=(F(t)\rv{x}(t))+e(\rv{x}(t),t)d\rv{v}(t),\qquad eQe^T=g_{ij}(t).
\end{align}
The state model is assumed to be affine and linear and possibly explicitly time dependent. The diffusion matrix is assumed to be independent of state, but possibly (explicitly) time dependent. 

Observe that this includes the case studied in the previous section. Since the diffusion vielbein is state dependent (or the state model noise is multiplicative), the methods used in the previous section cannot be applied to this case. This model is also more general than the Yau filter case where no explicit time dependence is assumed.   

The FPKfe for the transition probability density $P(t,x|t',x')$ is 
\begin{align}
        {\left\lbrace\begin{aligned}    
		\frac{\partial P}{\partial t}(t,x|t_0,x_0)&=-\sum_{i,j=1}^n\frac{\partial}{\partial x_i}\left(\left( F_{ij}(t)x_{j}+l_i(t) \right)P(t,x|t',x')\right)+\frac{1}{2}\sum_{i,j=1}^ng_{ij}(t)\frac{\partial^2P}{\partial x_i\partial x_j}(t,x|t',x'), \\ 
        P(t,x|t,x_0)&=\delta^n(x-x_0).\end{aligned}
\right.}
\end{align}
Let $\tilde{P}$ be the Fourier transform of $P$ with respect to $x$, i.e., 
\begin{align}
	P(t,x|t_0,x_0)=\frac{1}{(2\pi)^n}\int e^{i\sum_{i=1}^nk_ix_i}\tilde{P}(t,k|t_0,x_0)d^nk.
\end{align}
Since
\begin{align}
	\frac{\partial P}{\partial t}(t,x|t_0,x_0)&=\frac{1}{(2\pi)^n}\int e^{i\sum_{i=1}^nk_ix_i}\frac{\partial \tilde{P}}{\partial t}(t,k|t_0,x_0)d^nk,\\ \nonumber
	x_jP(t,x|t_0,x_0)&=\frac{1}{(2\pi)^n}\int e^{i\sum_{i=1}^nk_ix_i}x_j\tilde{P}(t,k|t_0,x_0)d^nk,\\ \nonumber
	&=\frac{1}{(2\pi)^n}\int\left[ (-i)\frac{\partial}{\partial k_i}e^{i\sum_{i=1}^nk_ix_i} \right]\tilde{P}(t,k|t_0,x_0)d^nk,\\ \nonumber
	&=\frac{i}{(2\pi)^n}\int e^{i\sum_{i=1}^nk_ix_i}\frac{\partial\tilde{P}}{\partial k_j}(t,k|t_0,x_0)d^nk,\\ \nonumber
	\frac{\partial P}{\partial x_i}(t,x|t_0,x_0)&=\frac{i}{(2\pi)^n}\int e^{i\sum_{i=1}^nk_ix_i}k_i\tilde{P}(t,k|t_0,x_0),\\ \nonumber
	\frac{1}{2}\sum_{i,j=1}^ng_{ij}(t)\frac{\partial^2P}{\partial x_i\partial x_j}(t,x|t_0,x_0))&=-\frac{1}{(2\pi)^n}\int e^{i'\sum_{i'=1}^nk_{i'}x_{i'}}\frac{1}{2}\sum_{i,j=1}^ng_{ij}(t)k_ik_j\tilde{P}(t,k|t_0,x_0)d^nk,
\end{align}
it follows that
\begin{align}
        {\left\lbrace\begin{aligned}    
		\frac{\partial \tilde{P}}{\partial t}(t,k|t_0,x_0)&=-\left(\sum_{i,j=1}^nF_{ij}(t)k_i\frac{\partial}{\partial k_j}+\sum_{i=1}^nl_i(t)k_i\right)\tilde{P}(t,k|t',x')-\frac{1}{2}\sum_{i,j=1}^ng_{ij}(t)k_ik_j\tilde{P}(t,k|t',x'), \\ 
		\tilde{P}(t,k|t,x_0)&=e^{i\sum_{j=1}^nk_jx_{0_j}}.\end{aligned}
\right.}
\end{align}
Since $P(t,x|t',x')$ is Gaussian, so is $\tilde{P}(t,k|t_0,x_0)$  and we can choose the following ansatz for $\tilde{P}(t,k|t_0,x_0)$: 
\begin{align}
	\tilde{P}(t,k|t_0,x_0)=e^{-i\sum_{i=1}^nk_i\mu_i(t,t_0)-\frac{1}{2}\sum_{i,j=1}^nk_ik_j\sigma_{ij}(t,t_0)}.
\end{align}
Without loss of generality, $\Sigma_{ij}=\Sigma_{ji}$. Note that if there is no explicit time dependence, the argument of $\mu$ and $\Sigma$ is the difference between the times, i.e., $t-t_0$.

Therefore
\begin{align}
	\left( -i\sum_{i=1}^nk_i\dot{\mu}_i-\frac{1}{2}\sum_{i,j=1}^nk_ik_j\dot{\Sigma}_{ij}+i\sum_{i,j=1}^nF_{ij}(t)k_i\mu_j+\sum_{i=1}^nl_ik_i+\sum_{i,j,l=1}^nF_{ij}k_i\Sigma_{jl}k_l+\frac{1}{2}\sum_{i,j=1}^ng_{ij}(t)k_ik_j \right)=0.
\end{align}
This implies that $\mu_i$ and $\Sigma_{ij}$ obey the following ODEs:
\begin{align}
        {\left\lbrace\begin{aligned}    
		\dot{\mu}_i(t,t_0)&=\sum_{j=1}^nF_{ij}(t)\mu_j(t,t_0)+l_i(t,t_0), \\ 
        \mu_i(t_0,t_0)&=x(t_0),\end{aligned}
\right.}
\end{align}
and
\begin{align}
	 {\left\lbrace\begin{aligned} 
		 \dot{\Sigma}_{ij}(t,t_0)&=\sum_{l=1}^n\Sigma_{il}(t,t_0)F_{lj}(t)+\sum_{l=1}^nF_{il}(t)\Sigma_{lj}(t,t_0)+g_{ij}(t),\\
		 \Sigma_{ij}(t_0,t_0)&=\Sigma(t_0).\end{aligned}\right.}
\end{align}
From the discussion in the previous section, the solution of these first-order ODEs are
\begin{align}
	\mu_i(t,t_0)=U_{ij}(t,t')x(t_0)+\int_{t_0}^tU(t,t_0)l(t)dt,
\end{align}
and
\begin{align}
	\Sigma_{ij}(t,t_0)=U(t,t_0)\Sigma(t_0)U^T(t,t_0)+\int_{t_0}^{t}U(t,\tau)G(\tau)U^T(t,\tau)d\tau',
\end{align}
so that
\begin{align}
	P(t,x|t_0,x_0)=\frac{1}{\sqrt{(2\pi)^n\det \Sigma(t,t_0)}}\exp\left( -\frac{1}{2}\left[ x-\mu(t,t_0) \right]^T\left[ \Sigma^{-1}(t,t_0) \right]\left[ x-\mu(t,t_0) \right] \right).
\end{align}
Thus, the expression is the same as derived earlier even for the more general case (i.e., although constant diffusion vielbein implies constant diffusion matrix (for constant $Q$, constant difusion matrix does not imply constant diffusion vielbein).

\section{Practical Implementation}\label{sec:PractImpl}

In this section, some  practical implementational aspects are discussed. Specifically, a computationally efficient filtering algorithm based the results derived in the previous sections is presented in Section \ref{ssec:SGFalgorithm}. Some additional aspects are discussed in Section \ref{ssec:SGFRemarks}. 

It is important to note that the transition probability is given in terms of an exponential function. This implies that $P(t'', x''|t', x')$ is non-negligible only in a very small region, or the transition probability density tensor is sparse. The sparsity property is crucial for storage and computation speed.

\subsection{Sparse Kernel Grid Filtering Algorithm} \label{ssec:SGFalgorithm}
The implementation of the continous-discrete filtering solution exploiting the sparsity property of the transtition probability density, or kernel, is thus straightforward. 
\begin{enumerate}
	\item Precompute the transition probability density is given by ($t''>t'$)
\begin{align}
	&P(t'',x''|t',x')=\frac{1}{\sqrt{(2\pi)^n\det\Sigma(t'',t')}}\\ \nonumber
	&\qquad\times\exp\left( -\frac{1}{2}\sum_{i,j=1}^n\left[ x_i''-\mu_i(t'',t') \right]\left[ \Sigma^{-1}(t'',t') \right]_{ij}\left[ x_j''-\mu_j(t'',t') \right] \right).
\end{align}

\item Threshold the transition probability density and save it as a sparse tensor. That is,  zero the entries corresponding to exponent larger than a certain (user-specified) threshold.  

\item  At time $t_k$
\begin{enumerate}
\item The prediction step is accomplished by
\begin{align}
	p(t_k|Y(t_{k-1}))=\int P(t_k,x|t_{k-1},x')p(t_{k-1},x'|Y(t_{k-1}))\left\{ d^nx' \right\}.
\end{align}
Note that $p(t_0|Y(t_0))$ is simply the initial condition $p(t_0,x_0)$.
\item The measurement at time $t_k$ are incorporated in the correction step via
\begin{align}
	p(t_k,x|Y(t_k))=\frac{p(y(t_k)|x)p(t_k,x|Y(t_{k-1}))}{\int p(y(t_k)|\xi)p(t_k,\xi|Y(t_{k-1}))}\left\{ d^n\xi \right\}.
\end{align}
\end{enumerate}
\end{enumerate}

This algorithm is referred to as the sparse grid filtering (SGF) algorithm. The thresholding step is crucial for real-time implementation. 

Note that the precomputation step assumes that the grid spacing and time intervals between measurements are known. 

In many applications, the measurement model is described by an additive Gaussian noise model, i.e.,
\begin{align}
	\rv{y}(t_k)=h(\rv{x}(t_k),t_k)+\rv{w}(t_k),\qquad k=1,2,\ldots,\qquad t_k>t_0,
\end{align}
with $\rv{w}(t)\sim N(0,R(t))$. Then, at observation time $t_k$, the conditional density is given by
\begin{align}
	p(t_k,x|Y(t_k))=\frac{p(y(t_k)|x)p(t_k,x|Y(t_{k-1}))}{\int p(y(t_k)|\xi)p(t_k,\xi|Y(t_{k-1}))\left\{ d^n\xi \right\}},
\end{align}
where $p(y(t_k)|x)$ is given by
\begin{align}\label{eq:MeasGaussCorr}
        p(y(t_k)|x)=\frac{1}{\left( (2\pi)^m\det R(t_k) \right)^{1/2}}\exp\left\{ -\frac{1}{2}(y(t_k)-h(x(t_k),t_k))^T(R
(t_k))^{-1}(y(t_k)-h(x(t_k),t_k)) \right\},
\end{align}
and $p(t_k,x|Y(t_{k-1}))$ is given by
\begin{align}
	p(t_k,x|Y(t_{k-1}))=\int P(t_k,x|t_{k-1},x_{k-1})p(t_{k-1},x_{k-1}|Y(t_{k-1}))\left\{ d^nx_{k-1} \right\}.
\end{align}

\subsection{Additional Remarks}\label{ssec:SGFRemarks}

Observe also that the fundamental solution has a simple and clear physical interpretation. Specifically, when the signal model noise is small the transition probability is significant only near trajectories satisfying the noiseless equation. The noise variance quantifies the extent to which the state may deviate from the noiseless trajectory.

The following additional observations can be made :
\begin{enumerate}
	\item It is noted that some of the results derived in this paper are well-known and used in EKF. In particular, the result of Section \ref{sec:LangFundSol} is used to derive the discrete-time version of the continuous-time state model (see, for instance, \cite{YaakovBar-ShalomandX.R.LiandT.Kirubarajan2001}). 

		However, the use of these results is very different from that in EKF. Specifically, the EKF linearizes the nonlinear measurement model and propagates the conditional mean and variance, assuming the prior conditional probablility density is Gaussian. Thus, the EKF fails if the nonlinearity is not benign, or if the Gaussian approximation for the prior is not valid (e.g., multi-modal). In the grid-based filtering,  the standard deviation computed from the conditional probability density is a reliable measure of the filter performance. This is not the case of suboptimal methods like the EKF.

		In other words, \textit{the SGF uses the exact expression for the fundamental solution that is given by a Gaussian function, while the EKF make the unjustified approximation that the conditional probability density at any time is Gaussian}.
	 	\item The conditional probability density calculated at grid points is exact (ignoring roundff errors) at those grid points. So, an interpolated version of the fundamental solution at coarser grid is close to the actual value. This suggests that a practical way of verifying the validity of the approximation is to note if the variation in the statistics with grid spacing, such as the conditional mean, is minimal.

		Observe that the time step size may be arbitrary. This means that the optimal solution (in the sense of being the correct solution) can be computed using this using a grid of sufficiently small grid spacing. However, the conditional state estimation will not be as good for large time steps, a fundamental limitation due to large measurement sample interval. 
	
	\item In this paper, the correction step was carried out using a uniform grid. It is clear that a much faster approach for higher dimensional problems would be to use Monte Carlo integration methods.

	\item The prediction step computation can be speeded up considerably by restricting calculation only in areas with significant conditional probability mass (or more accurately, in the union of the region of significant probability mass of $p(y(t_k)|x)$ and $p(x|Y (t_{k-1})))$.
	\item Although it is possible to compute the transition probability density off-line, it may be more appropriate to compute it on-line computing only those terms where initial points are in the region where the density is significant. It is also then possible to use an `adaptive grid', both in resolution and location/domain. 	Observe that the one-step approximation of the path integral can be stored more compactly as it is Gaussian. Note that when the noise is small, grid spacing needs to be finer in order to adequately sample the conditional probability density. 
	  
	\item Although the exact analytical expression of the conditional probability density is unknown, the accuracy of the grid-based method can be made very high and robust with small enough grid spacing. This situation is similar to the case of numerical solution of partial differential equations, where the exact solution is not known, but high accuracy numerical solutions (even machine precision) can still be obtained.  Thus, for practical purposes ``numerically exact'' solutions obtained with a fine grid are optimal.  Note that in the case of the PDEs, even when the exact solution is known, it often does not yield accurate numerical solutions.

\end{enumerate}

\section{ Comments on Discrete-Discrete Sparse Kernel Grid Filtering and Particle Filtering}\label{sec:Disc-DiscFilt}

\subsection{Discrete-Discrete Filtering}
In Section \ref{sec:LangFundSol}, the continuous-time state process has been converted to a discrete-time state sequence. This result is simple and exact because the model is linear. This is not possible for a general nonlinear state model, i.e., there is no such general formula relating the continuous-time process with its equivalent discrete-time sequence. In fact, a simple, nonlinear state process is not equivalent to a simple nonlinear state sequence. 

Observe that a discrete-time sequence description is more general than a continuous-time process. This is because less is assumed in the specification of the state sequence (for example, no assumption on continuity, differentiability are needed). However, the fundamental dynamical laws of classical physics are expressed in terms of continuous-time ODEs. Note that a simple, nonlinear dynamical model may not result in a ``simple'' discrete-time dynamical model.   

Often, a nonlinear discrete-time state process, or sequence, is studied. This can be either as a simple (e.g., Euler) approximation of a continuous-time state model, or directly via other means. In this section, the state sequence model is assumed to be given. Thus, one is led to consider the following general discrete-discrete filtering problem where the signal and measurement sequences are as follows:
\begin{align}
        {\left\lbrace\begin{aligned}    
		\rv{x}_k&=f(\rv{x}_{k-1},\rv{v}_{k-1},k),\\ 
		\rv{y}_k&=h(\rv{x}_k,\rv{n}_k,k).
\end{aligned}
\right.}
\end{align}
Here, $f:\mathbb{R}^{n_x}\times\mathbb{R}^{n_v}\ra\mathbb{R}^{n_x}$ is a nonlinear function of the state $\rv{x}_{k-1},\left\{ \rv{v}_{k-1},k\in\mathbb{N} \right\}$ is an independent, identically distributed (i.i.d) process noise sequence, and $n_x,n_v$ are dimensions of the state and process noise vectors. Similarly, $h_k:\mathbb{R}^{n_x}\times\mathbb{R}^{n_n}\ra\mathbb{R}^{n_y}$ is a nonlinear function, $\left\{ \rv{n}_k,k\in\mathbb{N} \right\}$ is an i.i.d. measurement noise sequance, and $n_y,n_n$ are dimensions of the measurement and measurement noise vectors. 

The universal discrete-discrete filtering problem is to evolve the conditional probability  distribution of the state at time $k$, based on the set of all available measurements $\rv{y}_{1:k}$, i.e., $p(x_k|z_{1:k})$. The recursive solution is well-known and similar to that discussed in Section \ref{sec:RevCDFilt}. Thus, if $p(x_{k-1}|y_{1:k-1})$ is known (note $p(x_0|y_0)=p(x_0)$ is the prior, or initial pdf), then
\begin{align}
{\left\lbrace\begin{aligned}    
	p(x_k|y_{1:k-1})&=\int p(x_k|x_{k-1})p(x_{k-1}|y_{1:k-1})dx_{1:k-1},\qquad\text{(Prediction)}\\
	p(x_k|y_{1:k})&=\frac{p(y_k|x_k)p(x_k|y_{1:k-1})}{\int p(y_k|x_k)p(x_k|y_{1:k-1})dx_k },\qquad\text{(Correction)}.
\end{aligned}
\right.}
\end{align}
Note that $p(x_k|x_{k-1},y_{1:k-1})=p(x_k|x_{k-1})$ due to the first-order Markov property of the state process.

Observe that ``discrete'' here refers to the time, not the state; the states take values in the continuum ($\mathbb{R}^{n_x}$).

This solution assumes that $p(x_k|x_{k-1})$ and $p(y_k|x_k)$ are known. Of course, they are defined by the state and measurement processes. In a practical situation, a closed form expression for these quantities is desired. This is indeed possible for several cases of practical interest, such as additive Gaussian noise. 

Of course, no closed form expression for the conditional probability density is known for an arbitrary initial distribution. Also, such a closed form solution, if it exists, may be unsuitable for an accurate numerical evaluation.  

\subsection{Sparse Kernel Grid Filtering}

It is usually stated that grid-based methods are computationally prohibitive when dealing with high-dimensional spaces. However, the transition probability density tensor is sparse, with the sparsity determined by the grid spacing, grid size and signal model noise. Likewise, the correction due to measurements is going to be sparse. Then, the conditional probability density is significant only in a small fraction of the grid space. Therefore, the prediction  step needs to be carried out in a very small region of the grid space. This implies that the memory requirements and the number of flops is considerably less than that suggested by naive counting. 

The arguments in Section \ref{sec:PractImpl} carry over to this case as well. In particular, na\" ive flops and memory estimate are unduly pessimistic for excellent performance using SGF. 

\subsection{Some Remarks on Particle Filtering}

An alternative to grid based techniques is particle filtering (see, for instance, \cite{ArulampalamMaskellGordonClapp2002}). A recursive Bayesian filter is obtained by Monte Carlo simulations. The idea is to represent the required conditional probability density  by a set of random samples with associated weights and then to compute estimates based on these samples and weights. As the number of samples becomes very large, this  approaches the optimal Bayesian solution. It has been shown that the performance of a PF is very good for many cases. An important aspect of particle filtering is a good choice of the proposal density.  For a discrete-time state process with linear measurement process, optimal proposal density known for PF. Note that PF solution is not robust for the case when the measurement model is non-linear.  

A common problem with the particle filter is the degeneracy phenomenon, where after a few iterations, all but one particle will have negligible weight. In fact, the variance of the importance weights can only increase over time, and thus, it is impossible to avoid the degeneracy phenomenon. This degeneracy implies that a large computational effort is devoted to updating particles whose contribution  to the approximation to is almost zero. Methods of alleviating this problem include choosing a good importance density and resampling.

From our discussion, it follows that the sample requirements in PF are analogous to the number of grid points in SGF (which is set by the threshold). The key point is that the SGF method does not require evaluation at all grid points due to the sparsity of the conditional probability density and the transition probability density. 

It is also noted that as long as the grid spacing is small enough to adequately sample the conditional probability density, there are no problems in propagating the conditional probability density. 

The SGF approach is likely to be useful in providing an estimate of the number of particles needed for the particle filtering. It is an interesting exercise to compare the computational load of the SGF with that of a robust PF solution (which is problem dependent).

\section{Example}\label{sec:Examples}

In the examples,  use is made of the Tensor toolbox in MATLAB developed by Bader and Kolda \cite{SAND2006-7592}. It has the  multininear sparse tensor class, essential for SGF. 

\begin{figure}[!h]
        \begin{center}
                \scalebox{0.70}{\includegraphics{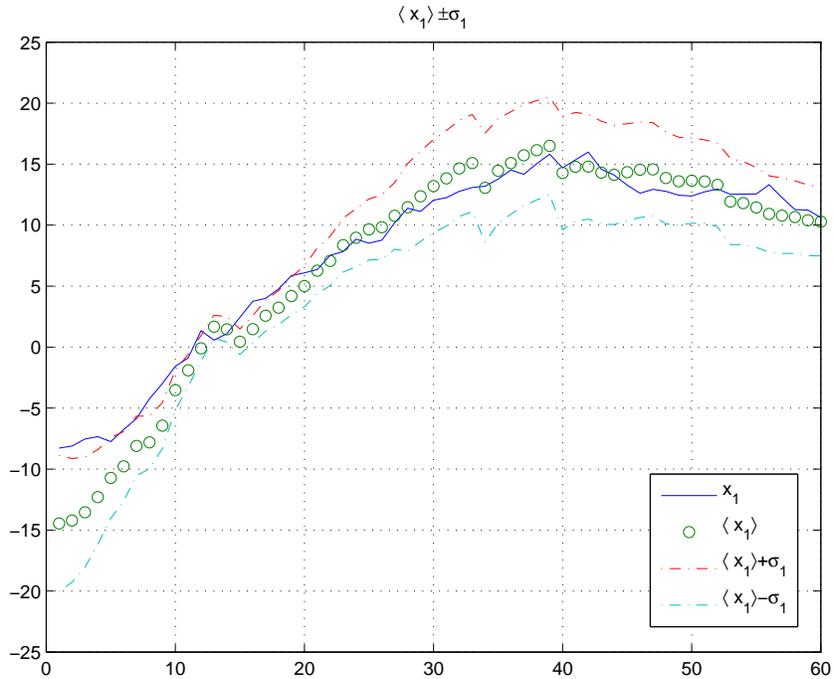}}
        \end{center}
	\caption{A sample of a the $\rv{x}_1$ state process, its conditional mean $\langle x_1\rangle$ and standard deviation $\sigma_1$. }
        \label{fig:YauASCMX1}
\end{figure}

Consider the 2D coordinated turn model with $\omega=\pi/30$ with $\hbar_{\nu}=1$ and time step size $T$ is 0.5 s. The measurement model drift is $h(x)=\tan^{-1}(x_2/x_1)$, or the angle,  with noise variance $(\pi/30 \text{\ rad})^2$. Thus, the measurement model is manifestly nonlinear function of the state variables. The transition probability density was computed in the interval $[-25,25]$ with grid spacing $\Delta x=1$ along $x_{1,2}$ using the Tensor toolbox in MATLAB \cite{ACM-TOMS-TENSORTOOLBOX}. Note that the Courant number $2\hbar_{\nu}T/\Delta x^2=1$  so that a one-step explicit scheme is unstable (an implicit scheme such as the Crank-Nicholson scheme is stable, but not as accurate  \cite{W.F.Ames1977}). 

Finally, the sparsity of the grid is about 99\% when the threshold is set to 10 (i.e., of exponent is greater than 10, the transition probability multilinear array element is set to be 0). The sparsity of $P(t'',x''|t',x')$ tensor is crucial in enabling a real-time computation of the $p(t,x|Y)$ on a current PC. In addition, it places considerably less memeory requirements. 

\begin{figure}[!h]
        \begin{center}
                \scalebox{0.70}{\includegraphics{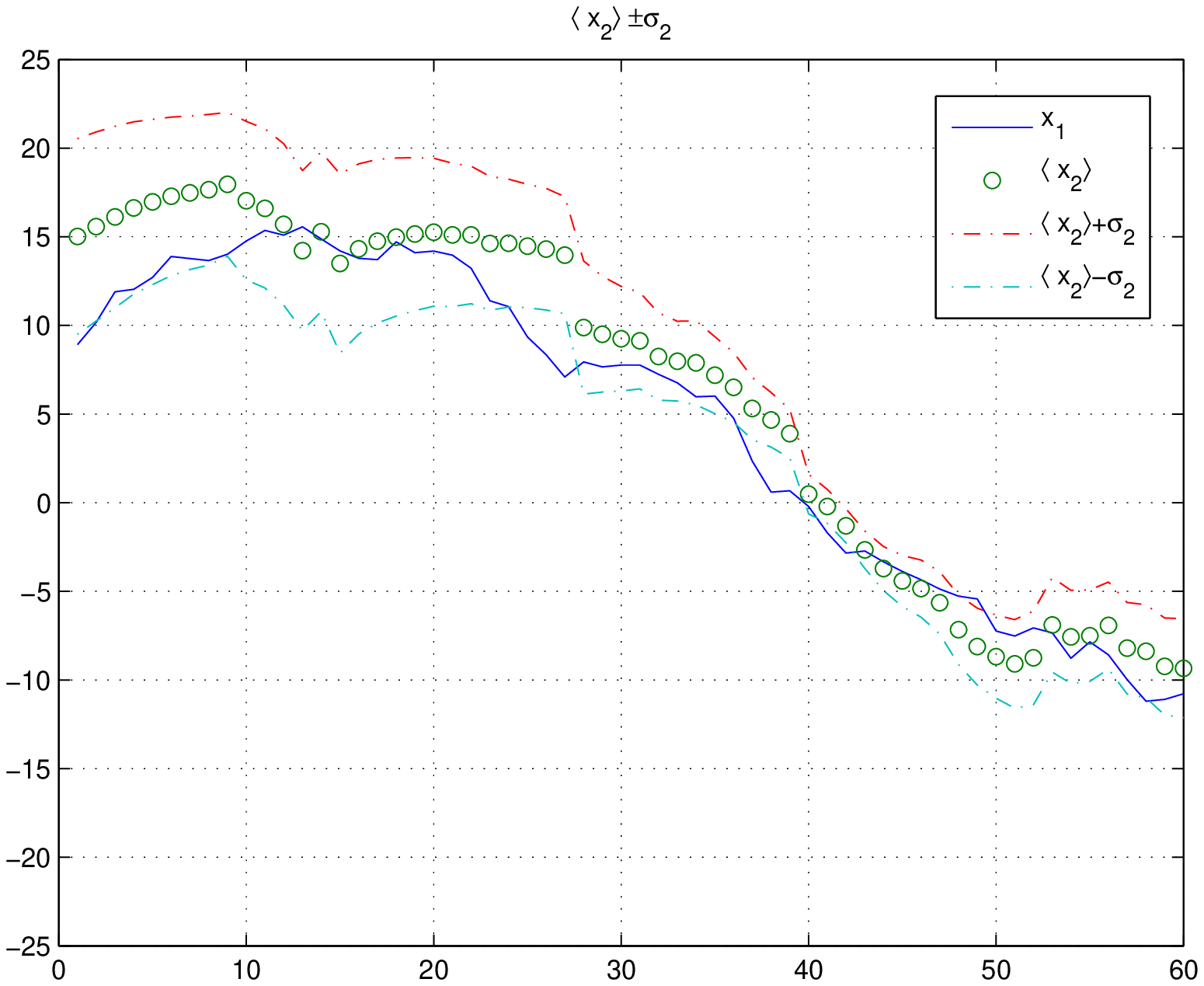}}
        \end{center}
	\caption{A sample of a the $\rv{x}_2$ state process, its conditional mean $\langle x_2\rangle$ and standard deviation $\sigma_2$. }
        \label{fig:YauASCMX2}
\end{figure}

Figures \ref{fig:YauASCMX1} and \ref{fig:YauASCMX2}  plot that the result obtained used the formula for the fundamental solution and using SGF. Also plotted are the conditional means and $\sigma$ bounds using the conditional probability density. It is seen that the tracking performance is very good in the sense that the conditional mean is found to be within a $\sigma$ most of the time. The root mean square errors for the two states over 100 Monte Carlo runs is found to be around 3 at $t=60$. Observe that the transition probability denisty tensor needed to be computed only once. 

In summary, proposed SGF method is seen to be stable and accurate even for such large time steps; the errors introuced in sprasifying the transition probability density are seen to be minimal, while the computational savings (in terms of speed and memory) are immense.

\section{Conclusion}

In this paper it has been shown that the continuous-discrete filtering problem with an affine, linear state model and with state independent diffusion matrix can be solved accurately using the exact fundamental solution of the corresponding FPKfe valid for an arbitrary time step size. Unlike the continuous-continuous case studied by Stephen Yau and collaborators, there are no restrictions on the measurement model in the continuous-discrete case. The sparsity of the transition probability density is then exploited to demonstrate that sparse grid filtering techniques can be used to solve higher dimensional problems with considerably less computational effort. Similar comments apply to the discrete-discrete filtering problem as well. 

It is also interesting to note that the path integral approach can be used to derive the expression for the fundamental solution for the FPKfe for the nonlinear state model case. Results for the additive and multiplicative noise cases are derived in \cite{PAPER1} and \cite{PAPER3}. In fact, it has been shown that the curdest approximation of the path integral formulas leads to very accurate results (see \cite{B.Balaji2007}). Thus, the sparse grid filtering approach yields a unified approach to tackling the general nonlinear filtering problem. Since it enables one to focus computation in regions of significant probability mass, it offers the possibility  of solving higher dimensional filtering problems in real-time. 

\appendix

\section{General Time Dependent Case}\label{sec:GenTimeDep}

The state transition matrix is the solution of the following equation:
\begin{align}\label{eq:GenTimDepModel}
 {\left\lbrace\begin{aligned}    
	 \frac{dU}{dt}(t,t_0)&=F(t)U(t,t_0),\\ 
	 U(t_0,t_0)&=I.
\end{aligned}
\right.}
\end{align}
The results derived in Section \ref{sec:LangFundSol} and \ref{sec:FundSolConsDiff} are not valid when the time-dependent matrix, $F(t)$, does not commute at different times (note that $[F(t),F(t)]=0$, i.e., it commutes at equal times). 

A formal solution for the general time-dependent case can be written down using the concept of time-ordered product. Let $F_i(t_i), i=1,\ldots, r$ be a string of time-dependent matrices and let $t_1\le t_2\le t_n$. The, the time-ordered product 
\begin{align}
	T(F_{i_1}(t_{i_1})F_{i_2}(t_{i_2})\cdots F_n(t_{i_n}))=F_1(t_1)F_2(t_2)\cdots F_n(t_n).
\end{align}

In other words, the time-ordered product of a string of operators (with time as its argument) is defined to be the string rearranged so that the eariler time operators  are to the right of the later ones. In general, there is an ambiguity when the operators do not commute at equal times. In the case of interest, $F_i(t)=F(t)$, so they commute at equal times. 

Observe that under the time-ordering symbol, everything commutes. Hence, a time derivative can be taken in the usual manner:
\begin{align}
	\frac{d}{dt}T\left(e^{\int_{t_0}^td\tau F(\tau)}\right)&=T\left( F(t)e^{\int_{t_0}^td\tau F(\tau)} \right),\\ \nonumber
	&=F(t)T\left( e^{\int_{t_0}^td\tau F(\tau)} \right).
\end{align}
Here the exponential of the matrix is defined via the usual power series expansion of the exponential function. In the second step, use is made of the fact that $t$ is the latest time, so that time-ordering puts it on the leftist. Since $t$ is the latest time, it can be taken out of the time-ordering:
\begin{align}
	\frac{d}{dt}T\left( e^{\int_{t_0}^td\tau F(\tau)} \right)=F(t)T\left( e^{\int_{t_0}^td\tau F(\tau)} \right). 
\end{align}
Thus, the solution of Equation \ref{eq:GenTimDepModel} is 
\begin{align}\label{eq:DysonFor}
	U(t,t_0)=T\left( e^{\int_{t_0}^td\tau F(\tau)} \right),\qquad t>t'.
\end{align}
The satisfaction of the boundary condition is obvious. Also, uniqueness of the solution follows because it is the solution of a first-order differential equation with a given initial value. 

An alternative form is the following:
\begin{align}
	U(t,t_0)=1+\int_{t_0}^tdt_1F(t_1)+\int_{t_0}^tdt_1\int_{t_0}^{t_1}dt_2F(t_1)F(t_2)+\int_{t_0}^tdt_1\int_{t_0}^{t_1}dt_2\int_{t_0}^{t_2}dt_3F(t_1)F(t_2)F(t_3)+\cdots.
\end{align}
The correctness of the formula follows from observing that when the right hand side is differentiated with respect to time, each term gives the previous one times $F(t)$. Observe that the various factors of $F(t)$ are time-ordered, i.e., matrices on later time are at the left. In fact, equality of this expression to Equation \ref{eq:DysonFor} follows from the following identity:
\begin{align}
	\int_{t_0}^tdt_1\int_{t_0}^{t_1}dt_2\cdots\int_{t_0}^{t_{n-1}}dt_nF(t_1)F(t_2)\cdots F(t_n)=\frac{1}{n!}\int_{t_0}^tdt_1dt_2\cdots dt_nT\left\{ F(t_1)F(t_2)\cdots F(t_n) \right\}.
\end{align}

Equation \ref{eq:DysonFor} is referred to as Dyson's formula and plays a central role in perturbative calculations in quantum mechanics and quantum field theory.

\section{Application: Maneuvering Target Tracking Signal Models}\label{sec:AppManTargets}

In this section, we summarize the results for many of the linear models that arise in maneuvering target tracking problems. For a nice, up-to-date review, the reader is referred to \cite{X.R.LiV.P.Jilkov2003}. Since the models are not time-independent, we can express results in terms of the time difference, or equivalently, set the initial time to $0$.

\subsection{Nilpotent or Orthogonal Matrix }

The matrix exponential function can be explicitly written in a closed form using the power series method for the following two cases:
\begin{itemize}
	\item Nilpotent $F$, i.e., $F^r=0$ for some positive integer $r$;
	\item Skew-symmetric $F$, i.e., $F^T=-F$.
\end{itemize}

The simplest example is the white noise acceleration model is 
\begin{align}
	\begin{bmatrix}
		\dot{\rv{x}}_1(t)\\
		\dot{\rv{x}}_2(t)
	\end{bmatrix}=
	\begin{bmatrix}
		0&1\\
		0&0
	\end{bmatrix}
	\begin{bmatrix}
		\rv{x}_1(t)\\
		\rv{x}_2(t)
	\end{bmatrix}+
	\begin{bmatrix}
		0\\ 
		1
	\end{bmatrix}\sigma\rv{\nu}(t).
\end{align}
It is easily seen that 
\begin{align}
	U(t,0)=
	\begin{bmatrix}
		1&t\\
		0&1
	\end{bmatrix},\qquad	
	\Sigma(t)-\Sigma(0)=\sigma^2
	\begin{bmatrix}
		t^3/3&t^2/2\\
		t^2/2&t
	\end{bmatrix}.
\end{align}
The state model for the Wiener process acceleration model is 
\begin{align}
	\begin{bmatrix}
		\dot{\rv{x}}_1(t)\\
		\dot{\rv{x}}_2(t)\\
		\dot{\rv{x}}_3(t)
	\end{bmatrix}=
	\begin{bmatrix}
		0&1&0\\
		0&0&1\\
		0&0&0
	\end{bmatrix}
	\begin{bmatrix}
		\rv{x}_1(t)\\
		\rv{x}_2(t)\\
		\rv{x}_3(t)
	\end{bmatrix}+
	\begin{bmatrix}
		0\\
		0\\
		1
	\end{bmatrix}\sigma\rv{\nu}(t).
\end{align}
and the corresponding results are
\begin{align}
	U(t,0)=
	\begin{bmatrix}
		1&t&\frac{t^2}{2}\\
		0&1&t\\
		0&0&1
	\end{bmatrix}, \qquad
	\Sigma(t)-\Sigma(0)=\sigma^2
	\begin{bmatrix}
		t^5/20&t^4/8&t^3/6\\
		t^4/8&t^3/3&t^2/2\\
		t^3/6&t^2/2&t
	\end{bmatrix}.
\end{align}
Another state model with a nilpotent drift matrix is
\begin{align}
	\begin{bmatrix}
		\dot{\rv{x}}_1(t)\\
		\dot{\rv{x}}_2(t)\\
		\dot{\rv{x}}_3(t)\\
		\dot{\rv{x}}_4(t)\\
		\dot{\rv{x}}_5(t)
	\end{bmatrix}=
	\begin{bmatrix}
		0&1&0&0&0\\
		0&0&0&0&0\\
		0&0&0&1&0\\
		0&0&0&0&0\\
		0&0&0&0&0
	\end{bmatrix}
	\begin{bmatrix}
		\rv{x}_1(t)\\
		\rv{x}_2(t)\\
		\rv{x}_3(t)\\
		\rv{x}_4(t)\\
		\rv{x}_5(t)
	\end{bmatrix}+
	\begin{bmatrix}
		0&0&0\\
		1&0&0\\
		0&0&0\\
		0&1&0\\
		0&0&1
	\end{bmatrix}
	\begin{bmatrix}
		\rv{\nu}_1(t)\\
		\rv{\nu}_2(t)\\
		\rv{\nu}_3(t)
	\end{bmatrix}.
\end{align}
which becomes the two-dimensional constant velocity model when the fifth state variable is ignored. In this case,
\begin{align}
	U(t,0)=
	\begin{bmatrix}
		1&t&0&0&0\\
		0&1&0&0&0\\
		0&0&1&t&0\\
		0&0&0&1&0\\
		0&0&0&0&1
	\end{bmatrix},\qquad
	\Sigma(t)-\Sigma(0)=
	\begin{bmatrix}
		t^3/3&t^2/2&0&0&0\\
		t^2/2&t&0&0&0\\
		0&0&t^3/2&0&0\\
		0&0&0&t&0\\
		0&0&0&0&t
	\end{bmatrix}.
\end{align}
All these examples correspond to the nilpotent case.

The coordinated turn state model is the simplest state model with skew-symmetric drift matrix:
\begin{align}
	\begin{bmatrix}
		\dot{\rv{x}}_1(t)\\
		\dot{\rv{x}}_2(t)
	\end{bmatrix}=
	\begin{bmatrix}
		0&-\omega\\
		\omega&0
	\end{bmatrix}
	\begin{bmatrix}
		\rv{x}_1(t)\\
		\rv{x}_2(t)
	\end{bmatrix}+
	\begin{bmatrix}
		\sigma_1\rv{\nu}_1(t)\\
		\sigma_2\rv{\nu}_2(t)
	\end{bmatrix}.
\end{align}
The results in this case are
\begin{align} 
        U(t,0)
&=\sum_{r=0}^{\infty}\frac{[(-\omega t)J]^{2r}}{(2r)!}+\sum_{r=0}^{\infty}\frac{[(-\omega t)J]^{2r+1}}{(
2r+1)!},\\ \nonumber
&=I_2\left[\sum_{r=0}^{\infty}(-1)^r\frac{(-\omega t)^{2r}}{(2r)!}\right]+J\left[\sum_{r=0}^{\infty}(-1)
^r\frac{(-\omega t)^{2r+1}}{(2r+1)!}\right],\\ \nonumber
&=I_{2}\cos\omega t+J\sin\omega t,\\ \nonumber
        &=\begin{bmatrix}
                \cos\omega t&\sin\omega t\\
                -\sin\omega t&\cos\omega t
        \end{bmatrix}.
\end{align}
and 
\begin{align}
	\Sigma(t)-\Sigma(0)=\int_{0}^t
	\begin{bmatrix}
		\sigma_1^2\cos^2(\omega t)+\sigma_2^2\sin^2(\omega t)&(\sigma_2^2-\sigma_1^2)\cos(\omega t)\sin(\omega t)\\
(\sigma_2^2-\sigma_1^2)\cos(\omega t)\sin(\omega t)&\sigma_2^2\cos^2(\omega t)+\sigma_1^2\sin^2(\omega t)
	\end{bmatrix}.
\end{align}

The three-dimesional skew-symmetric drift matrix leads to the nearly constant turn state model
	\begin{align}
	\begin{bmatrix}
		\dot{\rv{x}}_1(t)\\
		\dot{\rv{x}}_2(t)\\
		\dot{\rv{x}}_3(t)
	\end{bmatrix}=
	\begin{bmatrix}
		0&-\omega_3&\omega_2\\
		\omega_3&0&-\omega_1\\
		-\omega_2&\omega_1&0
	\end{bmatrix}
	\begin{bmatrix}
		\rv{x}_1(t)\\
		\rv{x}_2(t)\\
		\rv{x}_3(t)
	\end{bmatrix}+
	\begin{bmatrix}
		\sigma_1\rv{\nu}_1(t)\\
		\sigma_2\rv{\nu}_2(t)\\
		\sigma_3\rv{\nu}_3(t)
	\end{bmatrix}.
\end{align}
The exponential can be evaluated by using the Cayley-Hamilton theorem which states that a matrix satisfies its own charactersitic equation. The characteristic equation for $\Omega$ is 
\begin{align}
        \lambda^3+\omega^2\lambda=0,\qquad \omega^2=\omega_1^2+\omega_2^2+\omega_3^2.
\end{align}
This implies that
\begin{align} 
        \Omega^3=-\omega^2\Omega,\qquad \text{or }\quad
        \widehat{\Omega}^3=-\widehat{\Omega},\qquad\text{where }\qquad \widehat{\Omega}=\frac{\Omega}{\omega}.
\end{align}
Therefore,  all powers of $\Omega$ are related to the first and second powers. From this we see that
\begin{align}
        U(t,0)&=I_3+\frac{\Omega}{\omega}\left[ \omega t-\frac{(\omega t)^3}{3!}+\cdots \right]+\frac{\Omega^2}{\omega^2}\left[ \frac{(\omega t)^2}{2!}-\frac{(\omega t)^4}{4!}+\cdots \right],\\ \nonumber
        &=I_{3}-\Omega\frac{\sin\omega t}{\omega}+\Omega^2\frac{(1-\cos\omega t)}{\omega^2},\\ \nonumber
        &=I_3-\widehat{\Omega}\sin\omega t+\widehat{\Omega}^2(1-\cos\omega t).
\end{align}
This is known as Rodrigues' formula. Clearly, when $\omega=0$, $e^{-\Omega t}=I_3$. Since
\begin{align}
\Omega^2=
\begin{bmatrix}
        \omega_1^2-\omega^2&\omega_1\omega_2&\omega_1\omega_3\\
        \omega_2\omega_1&\omega_2^2-\omega^2&\omega_2\omega_3\\
        \omega_1\omega_3&\omega_2\omega_3&\omega_3^2-\omega^2
\end{bmatrix},
\end{align}
we finally obtain
\begin{align}
U(t,0)=   \begin{bmatrix}
                \cos\omega t+\widehat{\omega}_1^2(1-\cos\omega t)&\widehat{\omega}_3\sin\omega t+\widehat{\omega}_1\widehat{\omega}_2(1-\cos\omega t)&-\widehat{\omega}_2\sin\omega t+\widehat{\omega}_1\widehat{\omega}_3(1-\cos\omega t)\\
                -\widehat{\omega}_3\sin\omega t+\widehat{\omega}_2\widehat{\omega}_1(1-\cos\omega t)&\cos\omega t+\widehat{\omega}_2^2(1-\cos\omega t)&\widehat{\omega}_1\sin\omega t+\widehat{\omega}_2\widehat{\omega}_3(1-\cos\omega t)\\
                \widehat{\omega}_2\sin\omega t+\widehat{\omega}_3\widehat{\omega}_1(1-\cos\omega t)&-\widehat{\omega}_1\sin\omega t+\widehat{\omega}_3\widehat{\omega}_2(1-\cos\omega t)&\cos\omega t+\widehat{\omega}_3^2(1-\cos\omega)
        \end{bmatrix}, 
\end{align}
and 
\begin{align}
        b(t)&=-\int_{t'}^{t''}B(t)ldt,\\ \nonumber
        &=(A(t'')-A(t'))l,
\end{align}
where
\begin{align}
A(t)= \begin{bmatrix}
        \frac{\sin\omega t}{\omega}+\widehat{\omega}_1^2\left( 1+\frac{\sin\omega t}{\omega} \right)&-\widehat{\omega}_3\frac{\cos\omega t}{\omega}+\widehat{\omega}_1\widehat{\omega}_2\left( 1-\frac{\sin\omega t}{\omega} \right)&\widehat{\omega}_2\frac{\cos\omega t}{\omega}+\widehat{\omega}_1\widehat{\omega}_3\left( 1+\frac{\sin\omega t}{\omega} \right)\\
        -\widehat{\omega}_3\frac{\cos\omega t}{\omega}+\widehat{\omega}_2\widehat{\omega}_1\left( 1+\frac{\sin\omega t}{\omega} \right)&\frac{\sin\omega t}{\omega}+\widehat{\omega}_2^2\left( 1+\frac{\sin\omega t}{\omega} \right)&-\widehat{\omega}_1\frac{\cos\omega t}{\omega}+\widehat{\omega}_2\widehat{\omega}_3\left( 1+\frac{\sin\omega t}{\omega} \right)\\
        -\widehat{\omega}_2\frac{\cos\omega t}{\omega}+\widehat{\omega}_3\widehat{\omega}_1\left( 1+\frac{\sin\omega t}{\omega} \right)&\widehat{\omega}_1\frac{\cos\omega t}{\omega}+\widehat{\omega}_3\widehat{\omega}_2\left( 1+\frac{\sin\omega t}{\omega} \right)&\frac{\sin\omega t}{\omega}+\widehat{\omega}_3^2\left( 1+\frac{\sin\omega t}{\omega} \right)
        \end{bmatrix}.
\end{align}

\subsection{Other Cases}

For the general case, a systematic and elegant method for computing the exponential of an arbitrary matrix is presented in \cite{ChengYau1September1997}. A particularly attractive feature is that explicit formulas can be systematically and elegantly  derived for an (up to) arbitrary $4\times4$ matrix $F$ in terms of $F$ and its eigenvalues. It requires the determination of the minimial polynomial and the characteristic polynomial\footnote{Recall that any matrix satisfies its characteristic equation. The minimal polynomial, $M(x)$, is the  monic polynomial of least degree such that $M(F)=0$. The minimal polynomial always divides the characterisitic polynomial. Since $F_1F_2=0$ does not imply $F_1=0$ and/or $F_2=0$  for arbitrary matrices $F_{1}$ and $F_{2}$, the minimal polynomial needs to be determined by explicit matrix multiplication.}. Of course, this technique could also have been applied to the orthogonal and nilpotent $F$ cases studied in the previous section.  

The  constant turn model with known turn rate  is defined as
\begin{align}
	\begin{bmatrix}
		\dot{\rv{x}}_1(t)\\
		\dot{\rv{x}}_2(t)\\
		\dot{\rv{x}}_3(t)\\
		\dot{\rv{x}}_4(t)
	\end{bmatrix}=
	\begin{bmatrix}
		0&1&0&0\\
		0&0&0&-\omega\\
		0&0&0&1\\
		0&\omega&0&0
	\end{bmatrix}
	\begin{bmatrix}
		\rv{x}_1(t)\\
		\rv{x}_2(t)\\
		\rv{x}_3(t)\\
		\rv{x}_4(t)
	\end{bmatrix}+
	\begin{bmatrix}
		0&0\\
		1&0\\
		0&0\\
		0&1
	\end{bmatrix}
	\begin{bmatrix}
		\rv{\nu}_1(t)\\
		\rv{\nu}_2(t)
	\end{bmatrix}.
\end{align}
The eigenvalues of the drift matrix are $0,0,+i\omega, -i\omega$, and the minimal polynomial is $\lambda(\lambda-i\omega)(\lambda+i\omega)$ and it follows that 
\begin{align}
	U(t,0)=
	\begin{bmatrix}
		1&\frac{\sin\omega t}{\omega}&0&-\frac{1-\cos\omega t}{\omega}\\
		0&\cos\omega t&0&-\sin\omega t\\
		0&\frac{1-\cos\omega t}{\omega}&1&\frac{\sin\omega t}{\omega}\\
		0&\sin\omega t&0&\cos\omega t
	\end{bmatrix},\qquad
	\Sigma(t)-\Sigma(0)=
	\begin{bmatrix}
		\frac{2(\omega t-\sin\omega t)}{\omega^3}&\frac{1-\cos\omega t}{\omega^2}&0&\frac{\omega t-\sin\omega t}{\omega^2}\\
		\frac{1-\cos\omega t}{\omega^2}&T&-\frac{\omega t-\sin\omega t}{\omega^2}&0\\
		0&-\frac{\omega t-\sin\omega t}{\omega^2}&\frac{2(\omega t-\sin\omega t)}{\omega^3}&\frac{1-\cos\omega t}{\omega^2}\\
		\frac{\omega t-\sin\omega t}{\omega^2}&0&\frac{1-\cos\omega t}{\omega^2}&T
	\end{bmatrix}.
\end{align}
In the constant turn model with unknown turn rate, $\omega$ is usually described by a Wiener process model.  Note that the resulting model is nonlinear:
\begin{align}
	\begin{bmatrix}
		\dot{\rv{x}}_1(t)\\
		\dot{\rv{x}}_2(t)\\
		\dot{\rv{x}}_3(t)\\
		\dot{\rv{x}}_4(t)\\
		\dot{\rv{x}}_5(t)
	\end{bmatrix}=
	\begin{bmatrix}
		0&1&0&0&0\\
		0&0&0&-\rv{x}_5(t)&0\\
		0&0&0&1&0\\
		0&\rv{x}_5(t)&0&0&0\\
		0&0&0&0&0
	\end{bmatrix}
	\begin{bmatrix}
		\rv{x}_1(t)\\
		\rv{x}_2(t)\\
		\rv{x}_3(t)\\
		\rv{x}_4(t)\\
		\rv{x}_5(t)
	\end{bmatrix}+
	\begin{bmatrix}
		0&0&0\\
		1&0&0\\
		0&0&0\\
		0&1&0\\
		0&0&1
	\end{bmatrix}
	\begin{bmatrix}
		\rv{\nu}_1(t)\\
		\rv{\nu}_2(t)\\
		\rv{\nu}_3(t)
	\end{bmatrix}.
\end{align}
Note that the state model for the 2D essentially constant speed/nearly coordinated turn  modelis
\begin{align}
	\begin{bmatrix}
		\dot{\rv{x}}_1(t)\\
		\dot{\rv{x}}_2(t)\\
		\dot{\rv{x}}_3(t)\\
		\dot{\rv{x}}_4(t)
	\end{bmatrix}=
	\begin{bmatrix}
		0&0&1&0\\
		0&0&0&1\\
		0&0&0&-\omega\\
		0&0&\omega&0
	\end{bmatrix}
	\begin{bmatrix}
		\rv{x}_1(t)\\
		\rv{x}_2(t)\\
		\rv{x}_3(t)\\
		\rv{x}_4(t)
	\end{bmatrix}+
	\begin{bmatrix}
		0&0\\
		0&0\\
		1&0\\
		0&1
	\end{bmatrix}
	\begin{bmatrix}
		\sigma_1\rv{x}_1(t)\\
		\sigma_2\rv{x}_2(t)
	\end{bmatrix}.
\end{align}
which is related to the constant turn model with known turn rate via a permutation of state variables.

The Singer model assumes a zero-mean first-order Markov model for the acceleration:
\begin{align}
	\begin{bmatrix}
		\dot{\rv{x}}_1(t)\\
		\dot{\rv{x}}_2(t)\\
		\dot{\rv{x}}_3(t)
	\end{bmatrix}=
	\begin{bmatrix}
		0&1&0\\
		0&0&1\\
		0&0&-\alpha
	\end{bmatrix}
	\begin{bmatrix}
		\rv{x}_1(t)\\
		\rv{x}_2(t)\\
		\rv{x}_3(t)
	\end{bmatrix}+
	\begin{bmatrix}
		0\\
		0\\
		1
	\end{bmatrix}\sigma\rv{\nu}(t).
\end{align}
The minimal polynomial is the same as the characteristic polynomial. The exponential is\footnote{For simplicity, in the remainder of the section, only the expression for $U(t,0)$ will be written down; the expressions for $\mu$ and $\Sigma$ follow from it straightforwardly.} 
\begin{align}
	U(t,0)&=A_2(t)F^2+A_1(t)F+I_3,\\ \nonumber
	&=
	\begin{bmatrix}
		1&A_1(t)&A_2(t)\\
		0&1&A_1(t)-\alpha A_2(t)\\
		0&0&\alpha^2A_2(t)-\alpha A_1(t)+1
	\end{bmatrix},
\end{align}
where
\begin{align}
	A_2(t)&=\frac{1}{\alpha^2}\left( e^{-\alpha t}-1+\alpha t \right),\\ \nonumber
	A_1(t)&=t.
\end{align}
A modified verison of the Singer model is the mean-adaptive acceleration model
\begin{align}
	\begin{bmatrix}
		\dot{\rv{x}}_1(t)\\
		\dot{\rv{x}}_2(t)\\
		\dot{\rv{x}}_3(t)
	\end{bmatrix}=
	\begin{bmatrix}
		0&1&0\\
		0&0&1\\
		0&0&-\alpha
	\end{bmatrix}
	\begin{bmatrix}
		\rv{x}_1(t)\\
		\rv{x}_2(t)\\
		\rv{x}_3(t)
	\end{bmatrix}+
	\begin{bmatrix}
		0\\
		0\\
		\alpha
	\end{bmatrix}\tilde{a}(t)+
	\begin{bmatrix}
		0\\
		0\\
		1
	\end{bmatrix}\sigma\rv{\nu}(t).
\end{align}
which can be solved easily using the expression for $U(t,0)$.

The state model
\begin{align}
	\begin{bmatrix}
		\dot{\rv{x}}_1(t)\\
		\dot{\rv{x}}_2(t)\\
		\dot{\rv{x}}_3(t)
	\end{bmatrix}=
	\begin{bmatrix}
		0&1&0\\
		0&0&1\\
		0&-\omega^2&0
	\end{bmatrix}
	\begin{bmatrix}
		\rv{x}_1(t)\\
		\rv{x}_2(t)\\
		\rv{x}_3(t)
	\end{bmatrix}+
	\begin{bmatrix}
		0\\
		0\\
		1
	\end{bmatrix}\sigma\rv{\nu}(t)
\end{align}
is referred to as the second planar constant turn model. As the eigenvalues are distinct ($0,\pm\omega$), the minimal polynomial is the characteristic polynomial. In this case
\begin{align}
	U(t,0)&=\frac{F^2}{\omega^2}(1-\cos\omega t)+tF\sin\omega t+I_3,\\ \nonumber
	&=\begin{bmatrix}
		1&t\sin\omega t&\frac{1}{\omega^2}(1-\cos\omega t)\\
		0&\cos\omega t&t\sin\omega t\\
		0&-\omega^2t\sin\omega t&\cos\omega t
	\end{bmatrix}.
\end{align}

The state model
	\begin{align}
	\begin{bmatrix}
		\dot{\rv{x}}_1(t)\\
		\dot{\rv{x}}_2(t)\\
		\dot{\rv{x}}_3(t)
	\end{bmatrix}=
	\begin{bmatrix}
	0&1&0\\
	0&-\beta&1\\
	0&0&-\alpha
	\end{bmatrix}
	\begin{bmatrix}
		\rv{x}_1(t)\\
		\rv{x}_2(t)\\
		\rv{x}_3(t)
	\end{bmatrix}+
	\begin{bmatrix}
		0\\
		1\\
		0
	\end{bmatrix}u(t)+
	\begin{bmatrix}
		0\\
		0\\
		1
	\end{bmatrix}\rv{\nu}(t).
\end{align}
refers to the Markovian jump mean acceleration model.  Since the eigenvalues are all distinct ($0,-\alpha,-\beta$), minimal polynomial is characteristic polynomial. The exponential of $F$ is given by 
\begin{align}
	U(t,0)&=A_2(t)F^2+A_1(t)F+I_3,\\ \nonumber
	&=
	\begin{bmatrix}
		1&A_1(t)-\beta A_2(t)&A_2(t)\\
		0&1+\beta^2A_2(t)-\beta A_1(t)&A_1(t)-(\alpha+\beta)A_2(t)\\
		0&0&1-\alpha A_1(t)+\alpha^2A_2(t)
	\end{bmatrix},
\end{align}
where
\begin{align}
	A_1(t)&=\frac{1}{\alpha\beta(\beta-\alpha)}\left[ (\alpha^2-\beta^2)+\beta^2e^{-\alpha t}-\alpha^2e^{-\beta t} \right],\\ \nonumber
	A_2(t)&=\frac{1}{\alpha\beta(\beta-\alpha)}\left[ (\beta-\alpha)-\beta e^{-\alpha t}+\alpha e^{-\beta t} \right].
\end{align}

The planar variable turn model is
\begin{align}
	\begin{bmatrix}
		\dot{\rv{x}}_1(t)\\
		\dot{\rv{x}}_2(t)\\
		\dot{\rv{x}}_3(t)
	\end{bmatrix}=
	\begin{bmatrix}
		0&1&0\\
		0&0&1\\
		0&-\beta&-\alpha
	\end{bmatrix}
	\begin{bmatrix}
		\rv{x}_1(t)\\
		\rv{x}_2(t)\\
		\rv{x}_3(t)
	\end{bmatrix}+
	\begin{bmatrix}
		0\\
		0\\
		1
	\end{bmatrix}\rv{\nu}(t).
\end{align}
Assuming $(\alpha/2)\neq\beta$, the eigenvalues are $0,\alpha_1,\alpha_2$, where $\alpha_{1,2}=-\frac{\alpha}{2}\pm\sqrt{\frac{\alpha^2}{4}-\beta}$. Since the eigenvalues are distinct, the characteristic polynomial is the minimal polynomial and one obtains
\begin{align}
U(t,0)&=A_2(t)F^2+A_1(t)F+I_3,\\ \nonumber
&=
\begin{bmatrix}
	1&A_1(t)&A_2(t)\\
	0&1-\beta A_2(t)&-\alpha A_2(t)+A_1(t)\\
	0&\alpha\beta A_2(t)-\beta A_1(t)&(\alpha^2-\beta)A_2(t)-\alpha A_1(t)+1
\end{bmatrix},
\end{align}
where
\begin{align}
	A_1(t)&=\left[ \frac{(\alpha_1+\alpha_2)}{\alpha_1\alpha_2}+\frac{\alpha_2e^{-\alpha_1t}}{\alpha_1(\alpha_1-\alpha_2)}-\frac{\alpha_1e^{-\alpha_2t}}{\alpha_2(\alpha_1-\alpha_2)} \right],\\ \nonumber
	A_2(t)&=\left[ \frac{1}{\alpha_1\alpha_2}+\frac{e^{-\alpha_1t}}{\alpha_1(\alpha_1-\alpha_2)}-\frac{e^{-\alpha_2t}}{\alpha_2(\alpha_1-\alpha_2)} \right].
\end{align}

The state model for a certain class of oscillatory targets is
	\begin{align}
	\begin{bmatrix}
		\dot{\rv{x}}_1(t)\\
		\dot{\rv{x}}_2(t)\\
		\dot{\rv{x}}_3(t)
	\end{bmatrix}=
	\begin{bmatrix}
		0&1&0\\
		0&0&1\\
		-\gamma&-\beta&-\alpha
	\end{bmatrix}
	\begin{bmatrix}
		\rv{x}_1(t)\\
		\rv{x}_2(t)\\
		\rv{x}_3(t)
	\end{bmatrix}+
	\begin{bmatrix}
		0\\
		0\\
		1
	\end{bmatrix}\rv{\nu}(t).
\end{align}
Let the eigenvalues be $\lambda_1,\lambda_2,\lambda_2$. The expression for $U(t,0)$ is
\begin{align*}
	\frac{1}{(\lambda_1-\lambda_2)(\lambda_2-\lambda_3)(\lambda_3-\lambda_1)}&\Big[ (\lambda_2-\lambda_3)(F-\lambda_2I_3)(F-\lambda_3I_3)\\ \nonumber
	&\qquad+(\lambda_3-\lambda_1)(F-\lambda_3I_3)(F-\lambda_1I_3)+(\lambda_1-\lambda_2)(F-\lambda_1I_3)(F-\lambda_1I_3) \Big].
\end{align*}

The Markov acceleration model for constant turns 
\begin{align}
	\begin{bmatrix}
		\dot{\rv{x}}_1(t)\\
		\dot{\rv{x}}_2(t)\\
		\dot{\rv{x}}_3(t)\\
		\dot{\rv{x}}_4(t)
	\end{bmatrix}=
	\begin{bmatrix}
		0&1&0&0\\
		0&0&1&0\\
		0&0&0&1\\
		0&0&-\alpha_2&-\alpha_1
	\end{bmatrix}
	\begin{bmatrix}
		\rv{x}_1(t)\\
		\rv{x}_2(t)\\
		\rv{x}_3(t)\\
		\rv{x}_4(t)
	\end{bmatrix}+
	\begin{bmatrix}
		0\\
		0\\
		-\beta_1\\
		\beta_2-\alpha_1\beta_1
	\end{bmatrix}\rv{\nu}(t).
\end{align}
also arises for a class of oscillatory targets. The characteristic equation for $F$ is
\begin{align}
	\lambda^2\left[ \lambda(\lambda+\beta)+\alpha \right]=0,
\end{align}
so that the eigenvalues are $0,0,\alpha_1,\alpha_2$ where $\alpha_{1,2}=-\frac{\alpha}{2}\pm\sqrt{\frac{\alpha^2}{4}-\beta}$. The minimal polynomial is found to be identical to the characteristic polynomial. From a theorem in  \cite{ChengYau1September1997} , one obtains
\begin{align}
	U(t,0)=(1+tF)+F^2\left[ f_1(t)(F-\alpha_1)+f_2(t)(F-\alpha_2) \right],
\end{align}
where
\begin{align}
	f_1(t)&=\frac{e^{\alpha_1t}-(1+\alpha_1t)}{\alpha_1^2(\alpha_1-\alpha_2)},\\ \nonumber
	f_2(t)&=\frac{e^{\alpha_2t}-(1+\alpha_2t)}{\alpha_2^2(\alpha_2-\alpha_1)}.
\end{align}
\bibliographystyle{IEEEtran}
\bibliography{onfbib}

\end{document}